\documentclass[a4paper,12pt]{article}

\usepackage{graphics}

\begin{document}

\title{
\bf The inhomogeneous equation of state and the road towards the solution of the cosmological constant problem}

\author{Hrvoje \v Stefan\v ci\'c\thanks{shrvoje@thphys.irb.hr}
}

\vspace{3 cm}
\date{
\centering
Theoretical Physics Division, Rudjer Bo\v{s}kovi\'{c} Institute, \\
   P.O.Box 180, HR-10002 Zagreb, Croatia}


\maketitle

\abstract{We present a cosmological model containing a cosmological constant $\Lambda$ and a component with an inhomogeneous equation of state. We study the form of the inhomogeneous equation of state for which the model exhibits the relaxation of the cosmological constant, i.e. it asymptotically tends to the de Sitter regime characterized by a small positive effective cosmological constant. The effect of the relaxation of the cosmological constant is observed both for negative and positive values of $\Lambda$ and for a range of model parameters. A special emphasis is put on the study of the details of the CC relaxation mechanism and the robustness of the mechanism to the variation of model parameters. It is found that within the studied model the effective cosmological constant at large scale factor values is small because the absolute value of the real cosmological constant is large.}

\vspace{2cm}

\section{Introduction}

\label{int}

The understanding of cosmology has undergone a phase of intensive development during the last decade. It has been largely propelled by the arrival of more precise and abundant observational data \cite{SN,WMAP,LSS}. The picture of universe implied by the new data, however, revealed that the unknown part of the composition of the universe is far greater that previously believed. It is interesting to notice that the progress achieved in a great deal consists in establishing the fact that we know much less about our universe than we thought before the advent of the new observational data. The term unknown composition of the universe should not be understood too literally. It might really be the case that the universe contains a new physical component (or several of them), but it is also possible that the perceived unknown composition is just a manifestation of additional dimesions or modifications of interactions such as gravity. 

One of the most striking consequences of the unknown part of the composition of the universe is its present accelerated expansion which started at redshift values of the order 1. The accelerated expansion of the present universe is strongly confirmed by the observational data. Presently available data, however, provide much less information on the  actual cause of the acceleration. As already stated, the acceleration might happen owing to the existence of an unknown component with the negative pressure, referred to as {\em dark energy (DE)}, or might be a consequence of the fact that our universe has a number of macroscopic dimensions different than 4 or that the laws of gravitational interaction are modified at cosmological scales. These possibilities represent some of the most studied options leading to the accelerated expansion, but they certainly do not exhaust the list of proposals for the explanation of the accelerated expansion. The concept of dark energy seems especially useful in the modelling of the acceleration mechanisms. Namely, even if the acceleration is not due to some physical component with a negative pressure, the framework of dark energy can be used as a very good effective description of the alternative acceleration mechanisms which especially facilitates the comparisons of different approaches to the explanation of the accelerated expansion of the universe. 

It is also important to stress that the unknown composition of the universe is not entirely connected to the accelerated expansion of the universe. The observational data imply that our universe contains a significant component of nonrelativistic matter, also called {\em dark matter (DM)} which is important for the explanation of the growth and formation of the structures such as galaxies and clusters of galaxies that we observe in the universe today. The nature of dark matter also has not yet been firmly established. 

The dark energy component is primarily characterized by its negative pressure. A very large number of DE models have been proposed lately, of various degree of complexity, predictive potential and connection to fundamental physical theories \cite{Rev}. The observational data still provide a lot of space for dynamical DE models, but the central place of the allowed parametric space is occupied by a very simple DE model, a so called {\em $\Lambda$CDM model}. This model assumes that the DE component is actually a small positive cosmological constant (CC). The cosmological constant is a well known concept present in the theory of general relativity (GR) from the very first years of the development of the theory \cite{Wein,Stra,Nob}. An important observation is that the GR allows the existence of the CC, but it does not determine its size. So, in any cosmological model based on GR we are not concerned with the question whether the CC should be there or not, but with the problem of its size and sign. Since, therefore, the CC should already be an ingredient of cosmological model based on GR, it is very convenient to use such an object as a source of the acceleration of the universe. Indeed, an assumption of an existence of a small positive CC, together with the existence of dark matter, fits the observational data very well. From the observational side things work well: we have a model with few parameters that uses familiar concepts and fits the data well. 

Fundamental quantum physical theories, however, predict various contributions to the observed value of the cosmological constant. In quantum field theory (QFT) there exist very large zero-point energy contributions for each of the quantum fields. There are also contributions from condensates such as Higgs condensate or QCD condensates. These contributions should be added to, in principle arbitrary value of CC allowed in GR. Any attempt of calculation of contributions to the cosmological constant reveals that their size is many orders of magnitude larger than the observed value of the CC. The number of orders of magnitude differs with the choice of effective QFT cutoff, but in any case we have differences which raise a lot of suspicion, to put it mildly. But, in the end, it is not that individual contributions matter, but their sum. In principle, for the classical contribution to the CC we can choose the needed value and reproduce the observed value of $\Lambda$. The problem is that all contributions have to cancel to very many decimal places for this mechanism to be effective. This problem, referred to as {\em fine-tuning} plagues this explanation of the observed value of the cosmological constant. It is sometimes also called ``the old CC problem".

If theoretical considerations reveal such difficulties for the cosmological constant as a DE candidate, maybe we should opt for some of dynamical DE models. Even should the future observational data prove that dark energy (as a true component or as an effective representation of some other acceleration mechanism) is dynamical, it only relegates the CC problem to another level. Then we have to understand why the size of the CC is much smaller that the observed DE energy density or, possibly, why it is zero. The possibility that the CC is exactly zero would open the way to the solution of the CC problem by invoking some new symmetry. However, presently there is no proof for the existence of such a symmetry.

Therefore, the CC problem is difficult and it goes even beyond the issue of dark energy. Indeed, apart from the drastic problem of the size of the observed CC, there is another problem related to the cosmological constant. Namely, presently available observational data imply that the energy densities of nonrelativistic matter and the cosmological constant are of the same order of magnitude at present epoch of the expansion of the universe. The energy density of nonrelativistic matter scales very differently with the expansion than the CC. Therefore it is quite peculiar that these energy densities that have been  very different in size in the past and will presumably be very different in the future (in the context of the $\Lambda$CDM model), are presently of comparable size. This problem is also called ``coincidence problem". In this paper we shall mainly deal with ``the old CC problem" whereas the ``coincidence problem" will be only remotely commented. 

This paper further elaborates the mechanism of the relaxation of the cosmological constant proposed in \cite{ccrelax}. The relaxation of the cosmological constant is defined as a dynamical solution of the ``old CC problem" without the fine-tuning of the parameters of the model. Essentially, in \cite{ccrelax} we model the dark energy sector and study the asymptotic behavior at large scale factor values. The relaxation of the cosmological constant corresponds to the asymptotic de Sitter regime with a small positive effective cosmological constant. The mechanism is implemented in the framework of a cosmic component with an inhomogeneous equation of state \cite{Odin1}. In this paper we expand the model of \cite{ccrelax}, study some of its limitations and examine the robustness of the CC relaxation mechanism.

\section{The cosmological constant relaxation model}

\label{setup}

We consider a two component cosmological model containing a cosmological constant with the energy density $\rho_{\Lambda}$ and an additional cosmological component with the energy density $\rho$. It is assumed that the universe is spatially flat, $k=0$. The expansion of the universe is defined by the Friedmann equation  

\begin{equation}
\label{eq:H2}
H^2=\frac{8 \pi G}{3} (\rho_{\Lambda}+\rho) \, .
\end{equation}

The evolution of the second component with the expansion of the universe is defined by the standard equation of continuity

\begin{equation}
\label{eq:cont}
d \rho = -3 (\rho+p) \frac{da}{a} \, ,
\end{equation}
where its equation of state (EOS) has a nonstandard form 

\begin{equation}
\label{eq:p}
p=w \rho - 3 \zeta_0 (H^2+\beta)^{\alpha} \, .
\end{equation}
Here we take $\zeta_0>0$ and $\alpha$ is an arbitrary real parameter. The form of EOS given in (\ref{eq:p}) fits into the framework of the inhomogeneous equation of state \cite{Odin1}. The concept of the inhomogeneous equation of state was also studied in \cite{inhom0,inhom1,inhom2,Bamba}. The role in the inhomogeneous DE equation of state in the process of structure formation  was examined in \cite{Mota}.  In the remainder of this paper we study how the inhomogeneous equation of state contributes to the relaxation of the cosmological constant. The reference \cite{Odin1} (see especially the Appendix) shows that a possible way to understand the inhomogeneous equation of state is as an effective description of the modified gravity theories or time-dependent nonlinear viscosity. 

The modified theories of gravity study the extension of GR as a possible source of the acceleration mechanism active at present cosmological era \cite{Odinmodgrav,Faraonimodgrav}. An example of $f(R)$ gravity, free of instabilities, \cite{Odinmodgrav,Faraonimodgrav,Odinstab} was presented in \cite{ccrelax} showing that the mechanism of the CC relaxation could be realized directly in $f(R)$ modified gravity theories. A systematic study of modified gravity theories consistent with the solar system precision gravity tests \cite{Odintheor} is needed to establish the robustness of the CC relaxation mechanisms within the modified gravity theories. 

The concept of bulk viscosity, as a dissipation mechanism of imperfect cosmic fluid consistent with the symmetries of the FRW universe, was used for the  study of various phenomena in cosmology \cite{Weinvisc,Zim,Gron}. A very interesting possibility is that the bulk viscosity could potentially account for the present acceleration of the universe without the presence of dark energy \cite{colistete,avelino}. Here we consider generalization of the phenomenon of bulk viscosity. Namely, $H$ is not a variable of state of the imperfect fluid and a general dependence of the fluid pressure on $H$ represents a step out of standard framework of bulk viscosity. A more appropriate name would be time-dependent or nonlinear viscosity. This generalized concept of viscosity, however, proves to be very useful in the study of the present accelerated cosmic expansion \cite{avelino2}, peculiar properties of dark energy, including the phenomenon of the CC boundary crossing \cite{debrev} as well as other interesting phenomena \cite{CVbrev2,CVbrev1}. Furthermore, it is of particular interest to investigate how the concept of viscosity combines with the concepts of braneworlds and modified gravity \cite{brevmg3,brevmg2,brevmg1,brevcross,brane2,brane1}.


The expressions (\ref{eq:H2}), (\ref{eq:cont}) and (\ref{eq:p}) can be easily combined to obtain a dynamical equation for the evolution of the Hubble function

\begin{equation}
\label{eq:dynH}
d H^2 + 3(1+w)\frac{da}{a} \left( H^2 - \frac{8 \pi G \rho_{\Lambda}}{3} - \frac{8 \pi G \zeta_0}{1+w} (H^2+\beta)^{\alpha} \right) =0\, .
\end{equation}
Further we scale all quantities of interest and introduce the following notation

\begin{equation}
\label{eq:notation}
h=(H/H_X)^2, \;\; s=a/a_X, \;\; \lambda=8 \pi G \rho_{\Lambda}/3 H_X^2, \;\;
\xi=8 \pi G \zeta_0 H_X^{2(\alpha-1)}/(1+w)\, , \;\;  b=\beta/H_X^2\, .  
\end{equation}
Here $H(a_X)=H_X$. it is important to state that the value of $a_X$ can in principle take any value. It is not intrinsically constrained within the present model. Using this notation we can rewrite (\ref{eq:dynH}) as 

\begin{equation}
\label{eq:dynH2}
s \frac{d h}{d s} + 3 (1+w) (h - \lambda - \xi (h+b)^{\alpha})=0 \, ,
\end{equation}
with $h(1)=1$ defining the initial condition.

A thorough analysis of the model for the case $b=0$ was performed in \cite{ccrelax} \footnote{Note that in \cite{ccrelax} it was convenient to choose the parameter $\alpha$ somewhat differently that in the present paper.}. There it was shown that the CC relaxation mechanism was active for $\alpha < 0$. Here we proceed with the analysis of the same interval for $\alpha$ and examine the effects of the nonzero values of the parameter $b$. 

Next we consider the value $\alpha=-1$ as a representative and an analytically tractable case. The equation (\ref{eq:dynH2}) now reads 

\begin{equation}
\label{eq:alpha-1}
\frac{(h+b) \, d h}{(h-h_{*1})(h-h_{*2})} = -3 (1+w) \frac{d s}{s} \, .
\end{equation}
Here $h_{*1}$ and $h_{*2}$ stand for the zeros of the denominator of the expression at the left hand side of (\ref{eq:alpha-1}). Their respective values are given by the following expressions:

\begin{equation}
\label{eq:hstar1}
h_{*1}= \frac{1}{2} \left(\lambda-b+\sqrt{(\lambda+b)^2+4 \xi}\right)\,  \\
\end{equation}
and 
\begin{equation}
\label{eq:hstar2}
h_{*2}= \frac{1}{2} \left(\lambda-b-\sqrt{(\lambda+b)^2+4 \xi}\right) \, . 
\end{equation}

The differential equation (\ref{eq:alpha-1}) can be easily integrated and we arrive at the closed form solution for the dynamics of the scaled Huuble parameter with the scale factor:
 
\begin{equation}
\label{eq:solrelax}
\left( \frac{h-h_{*1}}{1-h_{*1}} \right)^{A_{1}} \left( \frac{h-h_{*2}}{1-h_{*2}} \right)^{A_{2}} = s^{-3(1+w)} \, ,
\end{equation}
Here $A_1=(b+h_{*1})/(h_{*1}-h_{*2})$ and  $A_2=-(b+h_{*2})/(h_{*1}-h_{*2})$. 

Before the analysis of the results presented above, we make a short summary of the results of paper \cite{ccrelax} which correspond to a specific value $b=0$. This is a starting point of our analysis since in this paper we are interested in how the nonvanishing value of parameter $b$ modifies the CC relaxation mechanism observed in \cite{ccrelax}. The dynamics of the Hubble function $h$ depends on all model parameters $\alpha$, $\lambda$, $\xi$ and $w$. 

\begin{figure}
\centerline{\resizebox{0.7\textwidth}{!}{\includegraphics{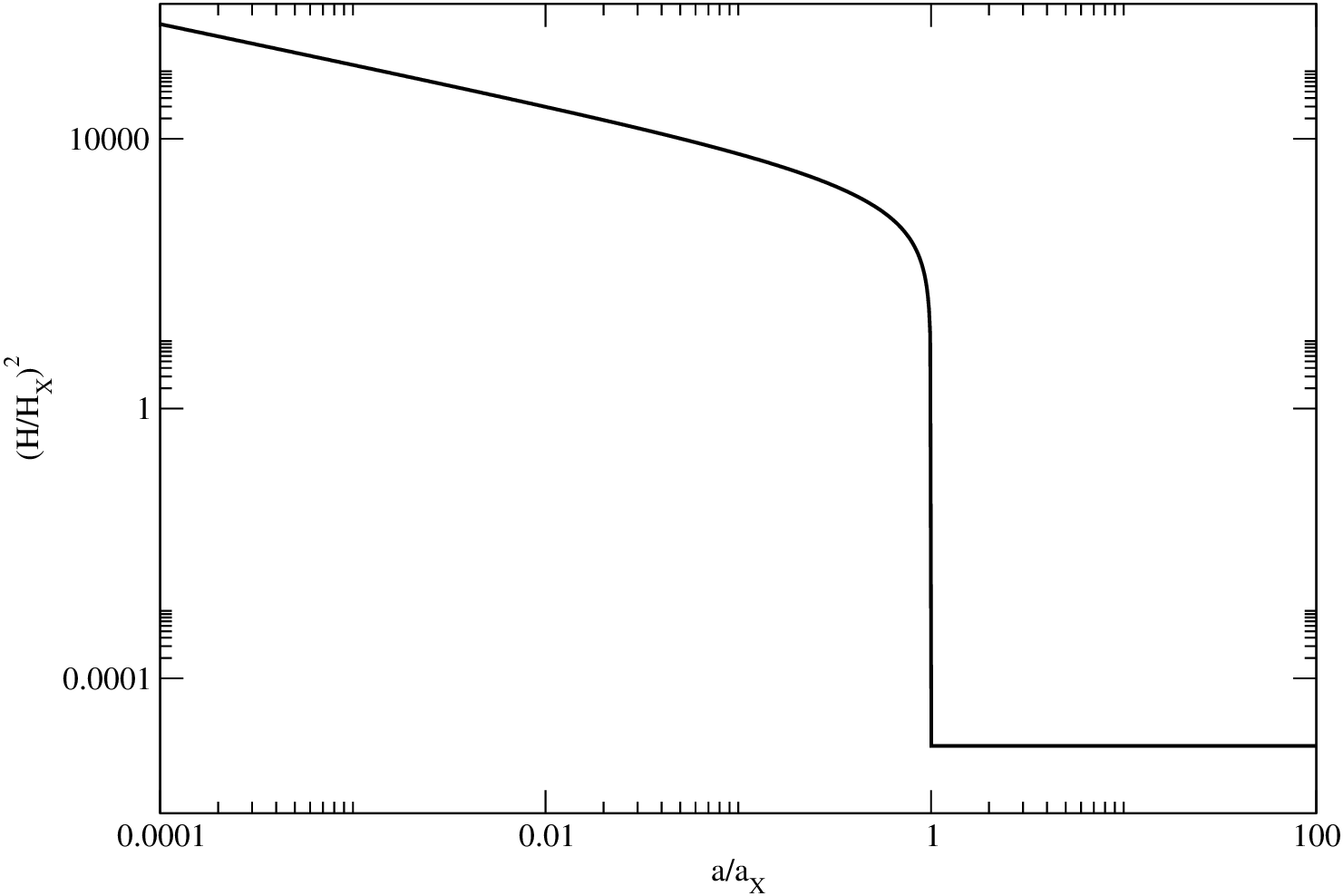}}}
\caption{\label{fig:brev1} The dependence of $h$ on the scale factor $a$ for the negative cosmological constant in the regime $\alpha < 0$. The parameter values used are $b=0$, $\alpha=-1$, $\lambda=-2000$, $\xi=0.02$ and $w=-0.8$.
}
\end{figure}

The Hubble function $h$ as a function of the scale factor for a case of negative $\lambda$ with a large absolute value and other representative parameter values ($\xi>0$, $\alpha<0$ and $w>-1$) is given in Fig. \ref{fig:brev1}. For these intervals of parameters the dynamics of $h$ is characterized by a very abrupt transition between a phase of expansion at small scale factor values where $h \sim a^{-3(1+w)}$ and a de Sitter phase at large scale factor values characterized by a small effective positive CC with $h_{asym} \sim \Lambda_{eff} \sim \xi/|\lambda|$. With other parameters fixed, the scale of $h$ before the transition grows with $|\lambda|$ and the value of $\Lambda_{eff}$ decreases with $|\lambda|$. The dynamics of $h$ before the transition is not affected by the size of $\xi$ whereas its asymptotic value at large $a$ grows with $\xi$. The choice of exponent $\alpha$ does not affect the behavior at small values of the scale factor, whereas the asymptotic value of $h$ at large $a$ decreases as $\alpha$ becomes more negative. Finally, as already stated, the value of $w$ affects the behavior before the transition and the large $a$ behavior of $h$ does not depend on $w$. For other choices of parameters the model may exhibit other interesting types of dynamics which however do not correspond to the CC relaxation mechanism.

\begin{figure}
\centerline{\resizebox{0.7\textwidth}{!}{\includegraphics{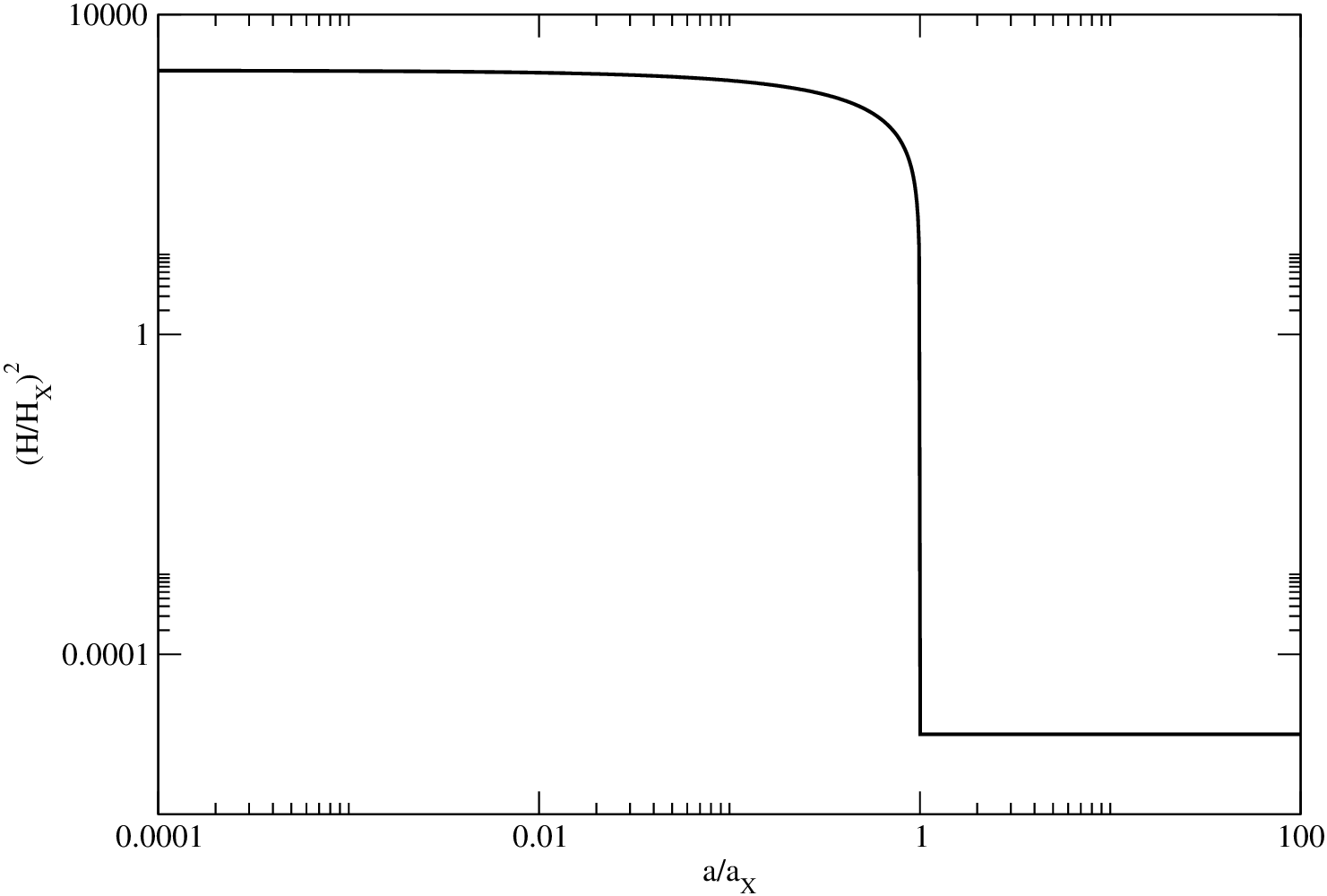}}}
\caption{\label{fig:brev2} The dynamics of $h$ as a function of the scale factor $a$ for a positive $\lambda$ and $\alpha<0$. The values of the used parameters are $b=0$, $\alpha=-1$, $\lambda=2000$, $\xi=-0.02$ and $w=-1.2$.
}
\end{figure}

The dynamics of the Hubble function $h$ for a large positive $\lambda$ and representative parameter values ($\xi<0$, $\alpha<0$ and $w<-1$) is given in Fig. \ref{fig:brev2}. In this regime, both for small and large scale factor values we find de Sitter regimes, $h \sim \lambda$ at small $a$ and $h \sim -\xi/\lambda$ at large $a$. These asymptotic regimes are again interconnected by an abrupt transition. The scale of the de Sitter regime preceding the transition grows and the scale of de Sitter regime following the transition decreases with the size of $\lambda$. The value of $\alpha$ does not affect the behavior at small $a$ values and the scale of the de Sitter regime at large $a$ grows as $\alpha$ becomes more negative. The asymptotic value of $h$ grows with the size of $|\xi|$ at large $a$ whereas the dynamics at small $a$ is not sensitive to the value of $\xi$. The value of parameter $w$ does not affect the dynamics of $h$ at large $a$ whereas the approach to de Sitter regime at small $a$ is sensitive to $w$. As in the case of negative $\lambda$, for other parameter intervals the behavior of the model is different and the CC relaxation mechanism is not effective. It is also important to stress that in the case of positive $\lambda$, the energy density $\rho$ should be negative. This is a strong argument to consider the second component as an effective description of some other fundamental mechanism.

Next we turn to the case of nonvanishing $b$. Since the values (\ref{eq:hstar1}) and (\ref{eq:hstar2}) determine the asymptotic behavior of the model, let us further study their dependence on the parameter $b$. We generally assume that the parameter $|\lambda|$ is by far the largest parameter of the model. More precisely, we suppose that $|\lambda|^2 \gg |\xi|$ and $|\lambda| \gg |b|$. These assumptions allow us to make an expansion of the square root terms in (\ref{eq:hstar1}) and (\ref{eq:hstar2}) with the following results:

\begin{equation}
\label{eq:h1exp}
h_{*1}=\frac{1}{2}(\lambda+|\lambda|)+\frac{b}{2}\left(\frac{|\lambda|}{\lambda}-1\right) + \frac{b^2+4 \xi}{4 |\lambda|} \, ,
\end{equation} 
\begin{equation}
\label{eq:h2exp}
h_{*2}=\frac{1}{2}(\lambda-|\lambda|)+\frac{b}{2}\left(-\frac{|\lambda|}{\lambda}-1\right) - \frac{b^2+4 \xi}{4 |\lambda|} \, .
\end{equation} 

The expressions differ for the cases of positive and negative $\lambda$. For $\lambda > 0$ we obtain

\begin{equation}
\label{eq:h1lpoz}
h_{*1}=\lambda + \frac{b^2+4 \xi}{4 |\lambda|} \, ,
\end{equation} 
\begin{equation}
\label{eq:h2lpoz}
h_{*2}=-b - \frac{b^2+4 \xi}{4 |\lambda|} \, .
\end{equation} 
On the other hand, for $\lambda < 0$ we have

\begin{equation}
\label{eq:h1lneg}
h_{*1}=-b + \frac{b^2+4 \xi}{4 |\lambda|} \, ,
\end{equation} 
\begin{equation}
\label{eq:h2lneg}
h_{*2}=\lambda - \frac{b^2+4 \xi}{4 |\lambda|} \, .
\end{equation} 
From the expressions (\ref{eq:h1lpoz}) to (\ref{eq:h2lneg}) we can determine the effect of the parameter $b$ on the asymptotic values. 

For a positive $\lambda$, at very small values of $|b|$ when $|b| \ll |\xi/\lambda|$ we have $h_{*1} \simeq \lambda$ and $h_{*2} \simeq -\xi/\lambda$. On the other hand, for a sufficiently large $|b|$, where $|b| \gg |\xi/\lambda|$, we have  $h_{*1} \simeq \lambda$ and $h_{*2} \simeq -b$. The asymptotic behavior at large scale factor values is determined by the value $h_{*2}$ and we see that at sufficiently large values of $|b|$, in the sense defined above, the parameter $b$ determines the asymptotic behavior of the Hubble function $h$. It is important to notice that in this case de Sitter regime at large values of the scale factor is realized only for negative values of $b$.

For negative values of $\lambda$ at small values for the parameter $b$, with $|b| \ll |\xi/\lambda|$ we obtain and $h_{*1} \simeq \xi/|\lambda|$ and $h_{*2} \simeq \lambda$. For $|b| \gg |\xi/\lambda|$ we further have $h_{*1} \simeq -b$ and $h_{*2} \simeq \lambda$. These results again show that for a sufficiently large value of $|b|$, this parameter determines $h_{*1}$ which in turn controls the asymptotic dynamics of the Hubble function at large scale factor values. Again, to have a de Sitter regime at large scale factor values $b$ has to be negative.   

A more careful analysis of the model immediately shows that for $h \rightarrow -b$ the pressure of the component with the inhomogeneous EOS diverges. The dynamics of the model reveals that this singular point is never reached. The expressions (\ref{eq:hstar1}) and (\ref{eq:hstar2}) show that the point where $p$ would diverge is never reached during the evolution of the model. For a large $|\lambda|$, the dynamics of $h$ stabilizes at a value slightly above $-b$.



\begin{figure}
\centerline{\resizebox{0.7\textwidth}{!}{\includegraphics{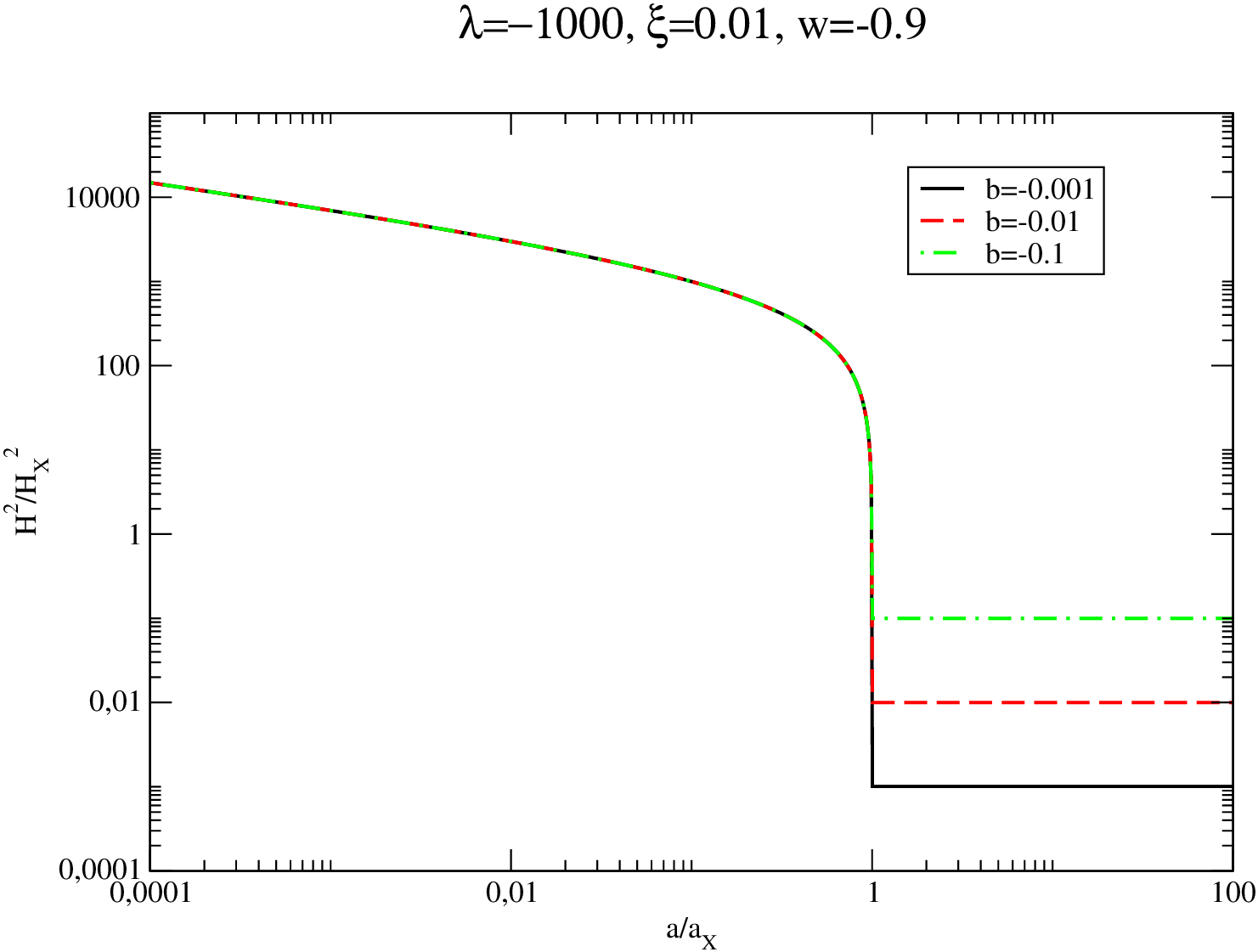}}}
\caption{\label{fig:1} The behavior of $h$ as a function of the scale factor $a$ for different values of the parameter $b$. The used parameter values are $\alpha=-1$, $\lambda=-1000$, $\xi=0.01$ and $w=-0.9$. In this parameter regime, the value of $b$ controls the asymptotic behavior of $h$ at large $a$.
}
\end{figure}

\begin{figure}
\centerline{\resizebox{0.7\textwidth}{!}{\includegraphics{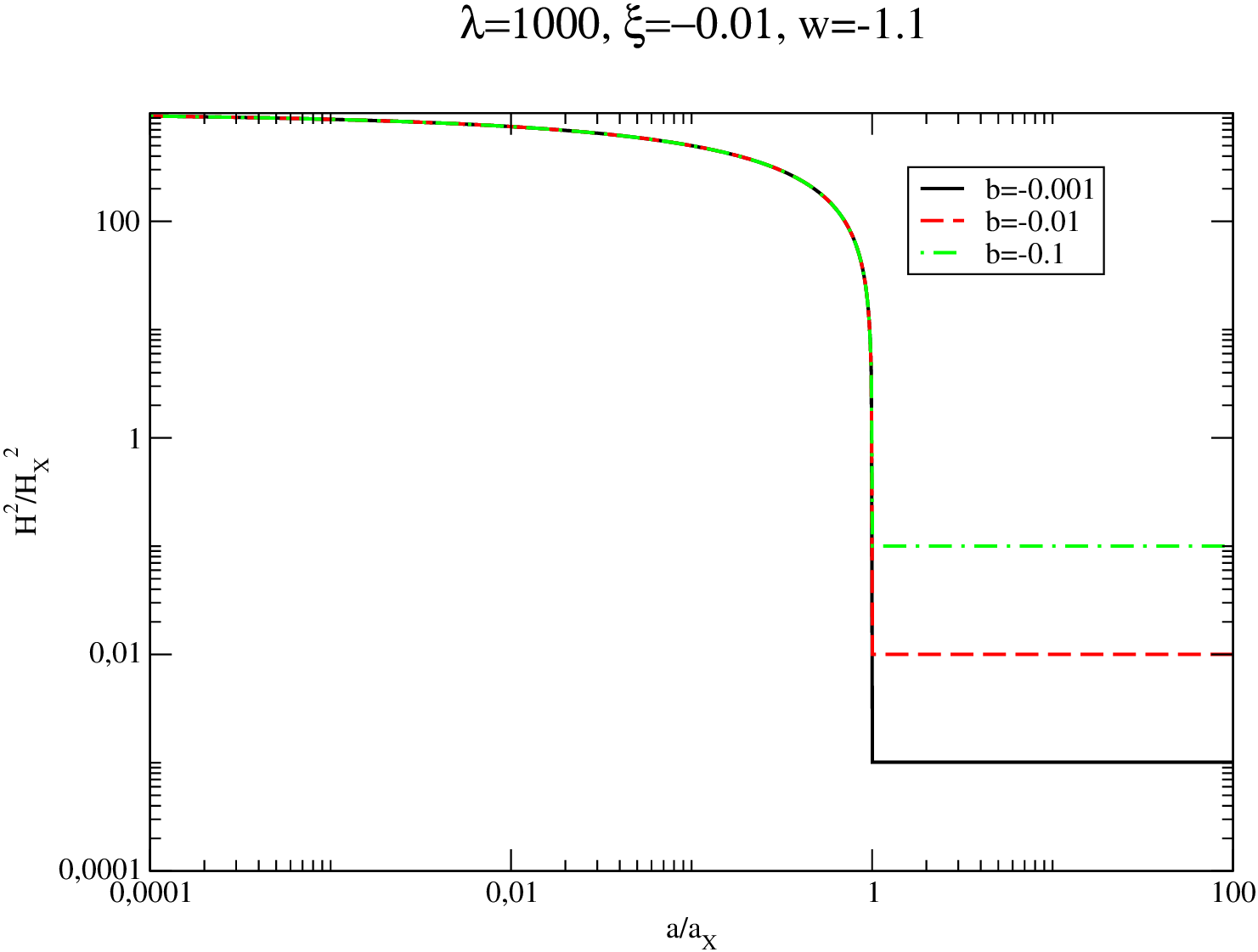}}}
\caption{\label{fig:2} The dependence of the dynamics of $h$ as a function of the scale factor on parameter $b$. The values of the parameters used are $\alpha=-1$, $\lambda=1000$, $\xi=-0.01$ and $w=-1.1$. The asymptotic behavior of $h$ at large $a$ is determined by the size of $b$.
}
\end{figure}

\section{Discussion}

The principal role of the considerations given above and the plots in Figures \ref{fig:1} and \ref{fig:2} is to gauge the role of the parameter $b$ in the model defined by (\ref{eq:H2})-(\ref{eq:p}). Our primary goal is to find out if and in which extent does the finite value of $b$ change the behavior of the model compared to the previously studied case corresponding to $b=0$. The qualitative behavior of the model retains the pattern observed for the vanishing value of $b$: both for positive and negative values of the scaled CC parameter $\lambda$ there is a distinct and abrupt transition from the expansion at a high energy density to the de Sitter regime. Therefore, this specific signature of the CC relaxation mechanism is not lost with the addition of the additional parameter $b$. The asymptotic value of $H^2$, however, depends on the interplay of all model  parameters. For a sufficiently small value of $b$, the asymptotic value is determined by the ratio of parameters $\xi$ and $\lambda$ ($h = |\xi/\lambda|$). 
As $b$ grows, the asymptotic value becomes fully dominated by the value of parameter $b$. 

Our aim is to look into a mechanism of the CC relaxation without fine-tuning. The case of vanishing $b$ possesses certain appeal since there the effective positive CC is small because $\lambda$ is large in absolute value and also the parameter $\xi$ is not expected to be large. For parameter values which we would expect based on the fundamental theories the expected value of the effective CC at large scale factor values is a small positive number. There is no need for the parameters to be fine-tuned.

For a large absolute value of $\lambda$, sufficiently large $b$ determines the asymptotic behavior of $H^2$. If we eventually aim at explaining the observed value of the cosmological constant, the value of $b$ should be small. Since there is no clear reason why the value of $b$ should be so small, some form of fine-tuning reenters into the model. We have to introduce a small parameter $b$ just to match the value of observed $\Lambda_{eff}$. Still, it is very important to stress that this value is very different from the real value $\Lambda$. Although a very large $\Lambda$ is present in the model, it does not determine decisively the asymptotic behavior of the system. Furthermore, the parameter $b$ plays the main role only because $\lambda$ is very large in absolute value. In a way the spirit of the CC relaxation mechanism is preserved: A universe with a large $\Lambda$ finally tends to a de Sitter state characterized by a small $\Lambda_{eff}$. The principal difference to the $b=0$ case is that for a sufficiently large $|b|$ there is no strong argument why $\Lambda_{eff}$ should be small.

The model of this paper represents and extension of the model studied in \cite{ccrelax}, but it is still just a starting point towards a realistic cosmological model with the resolved CC problem. Essentially, both the model of this paper and \cite{ccrelax} model the dark energy sector of the universe. Clearly, other components such as radiation and matter have to be added to create a realistic cosmological model and reproduce the standard eras of the evolution of the universe such as radiation dominated and matter dominated eras. The abruptness of the observed transition might pose a significant challenge to the construction of such a complete cosmological model. However, even in a model which contains the matter and radiation components, the presented CC relaxation mechanism should be efficient asymptotically since the energy densities of these components decay quickly with the expansion. Other important issues for further work are the timing of the transition, possible links to inflation and the growth of inhomogeneities in a universe in which the CC relaxation mechanism is active.

As discussed in the Introduction, there is another important problem related to the size of the dark energy (or CC) density. The coincidence problem is not directly addressed in the present model since the matter and radiation components are not present in the model. Only in the model with all relevant components this issue could be addressed properly.

Another very important issue is the physical motivation for the EOS given in (\ref{eq:p}). As already stated, the motivation could come from several directions of which we singled out modified gravity and generalized nonlinear viscosity. 
The elaboration of these topics might prove essential for a microscopic foundation of the mechanism exhibited by our model. 

\section{Conclusions}

The extension of the original model of the CC relaxation \cite{ccrelax} presented in this paper allows us the study of the limits of the CC relaxation mechanism. This is achieved through the introduction of the new parameter $b$. For a sufficiently large $|b|$, we no longer have an explanation of the smallness of $\Lambda_{eff}$ without fine-tuning. Still, even for larger values of $|b|$ the asymptotic value is not determined by a large $\lambda$, but some other small parameter, in particular $b$. In this sense, the spirit of the CC relaxation mechanism persists even for larger values of $|b|$. These conclusions show that the concept of the CC relaxation mechanism is robust, although for small or vanishing value of $b$ the solution of the CC problem is more natural. These findings further support the study of other types of inhomogeneous EOS as a road to a complete cosmological model in which the CC problem is naturally solved.

{\bf Acknowledgements.} 
The author would like to thank B. Guberina for useful comments on the manuscript. 
This work was supported by the Ministry of Education, Science and Sports of the Republic of Croatia 
under the contract No. 098-0982930-2864.

\end{document}